\documentclass[twocolumn,floatfix,superscriptaddress,amsmath,showpacs,showkeys,aps,prb]{revtex4}

\usepackage{t1enc}
\usepackage[final]{graphicx}
\usepackage{graphicx}
\usepackage{epsfig}
\usepackage{bm}

\usepackage{color}
\usepackage[normalem]{ulem}

\usepackage{graphics}

\begin{document}

\title{On the coexistence of dipolar frustration and criticality in ferromagnets}

\author{Sergio A. Cannas}
\email{cannas@famaf.unc.edu.ar}
\affiliation{Facultad de Matem\'atica, Astronom\'{\i}a, 
F\'{\i}sica y Computaci\'on, Universidad Nacional de C\'ordoba and \\ Instituto de
F\'{\i}sica Enrique Gaviola  (IFEG-CONICET), Ciudad Universitaria,
5000 C\'ordoba, Argentina}

\author{Alessandro Vindigni}
\email{vindigni@phys.ethz.ch}
\affiliation{$^1$Laboratorium f\"ur Festk\"orperphysik, Eidgen\"ossische Technische Hochschule Z\"urich, CH-8093 Z\"urich,
Switzerland}

\date{\today}

\begin{abstract}

In real magnets the tendency towards ferromagnetism -- promoted by exchange coupling --  is usually frustrated by dipolar interaction.
As a result, the uniformly ordered phase is replaced
by modulated (multi-domain) phases, characterized by
different order parameters rather than  the global magnetization. The transitions occurring within those modulated phases and towards the disordered phase are generally not of second-order type.  Nevertheless, strong experimental evidence indicates that a standard critical behavior is recovered when comparatively small fields are applied
that stabilize the uniform phase. The resulting power laws are observed with respect to a putative critical point
that falls in the portion of the phase diagram occupied by modulated phases, in line with an  avoided-criticality scenario.
Here we propose a generalization of the scaling hypothesis for ferromagnets, which explains this observation
assuming that the dipolar interaction acts as a relevant field, in the sense of renormalization group.
We corroborate this proposal with analytic and numerical calculations on the 2D Ising model frustrated by dipolar interaction.

\end{abstract}

\maketitle

\section{Introduction}
\label{Intro}

The continuous comparison with model experimental systems has played a crucial role in the development of the theory of cooperative phenomena.
In particular, magnetic systems have been a suitable playground for the study of second-order phase transitions.
In correspondence to a second-order phase transition observables follow a power-law behavior as a function of external parameters.
Such a power-law behavior defines the condition of \textit{criticality}. The property that different physical systems may follow the same power laws in the vicinity of the respective critical points 
is referred to as universality of critical exponents\cite{Taroni_2008,Stanley_RevMod_1999}.
Celebrated models that successfully reproduce this universal aspect of second-order phase transitions are (normally) based on short-ranged interactions\cite{Goldenfeld,LeBellac,Stanley_RevMod_1999}.
\begin{figure}[b!]
\includegraphics[width=8.5cm]{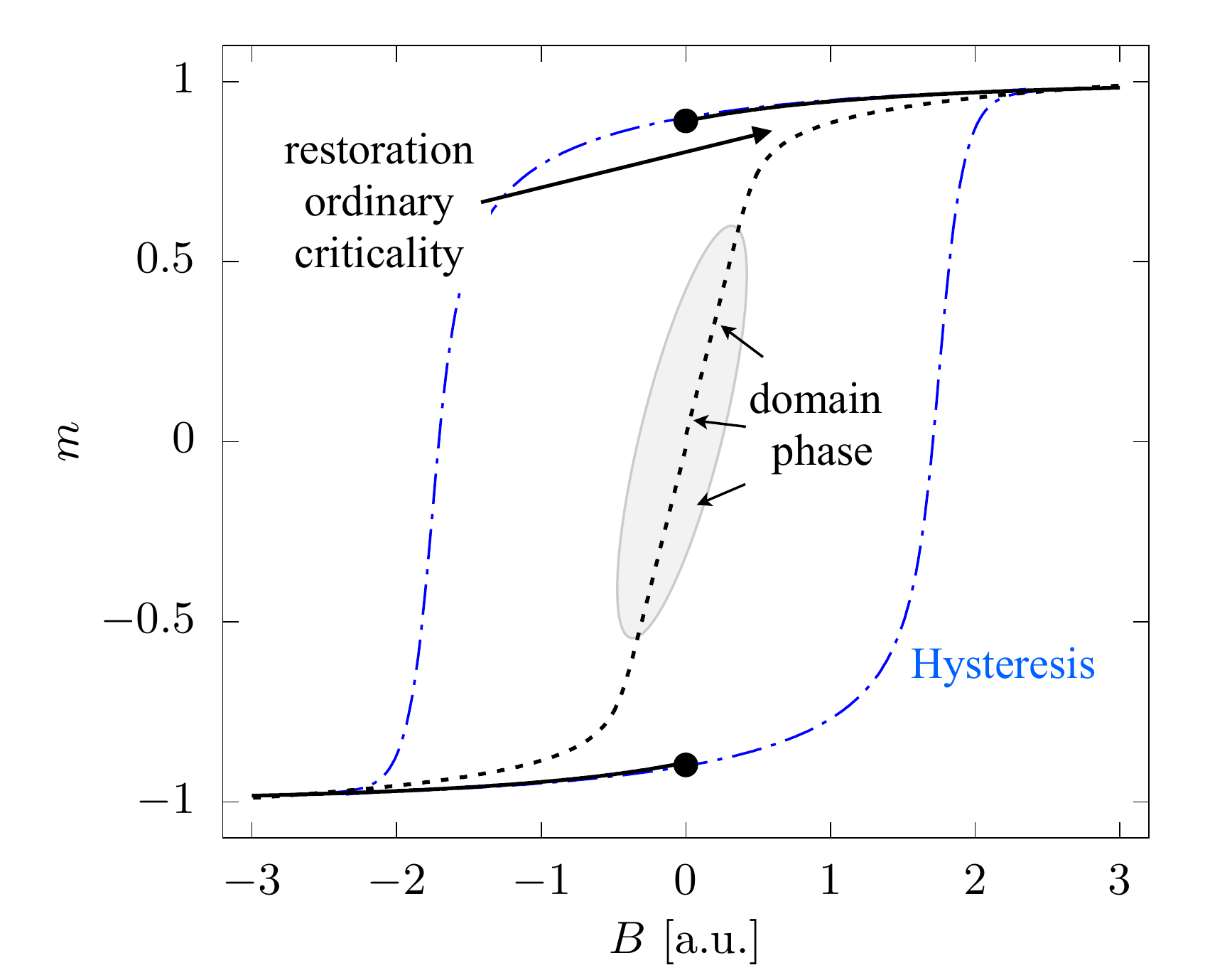}
\caption{\label{mag_curve}
Magnetization $m$ normalized to its saturation value as a function of $B$.
The solid and dashed lines represent the \textit{equilibrium} curves in the absence and in the presence of dipolar interaction, respectively.
The two bullets highlight the non-analyticity in $B=0$ expected only in the first case (see main text).
The ellipse highlights the portion of the dashed curve corresponding to the multi-domain phase, while the arrow indicates the region where standard criticality ($M\sim B^{1/\delta}$ for $T\simeq T_{\rm c}$) is restored.
The dot-dashed (blue) curve represents a prototypical hysteresis: both \textit{equilibrium} scenarios -- solid and dashed curves -- are compatible with hysteresis.
 }
\end{figure}
When applied to the ferromagnetic-to-paramagnetic phase transition, these textbook cooperative models are compatible with low-temperature
magnetization curves at thermodynamic \textit{equilibrium} similar to the discontinuous curve in Fig.~\ref{mag_curve}. The singularity in the magnetization curve originates from the very same non-analyticity that explains criticality and universality of critical exponents.
In practical cases, the magnetization as a function of the external field $B$ does not jump from one branch to the other when  $B=0$ is crossed:
the  system rather remains in a  \textit{metastable} configuration and the curve displays magnetic hysteresis, a typical \textit{out-of-equilibrium} phenomenon (Fig.~\ref{mag_curve}).
In real magnets the short-ranged exchange interaction, which drives the establishment of ferromagnetism at low temperature,
coexists with the long-ranged dipolar interaction.
This second interaction generally frustrates the realization of a phase with uniform magnetization throughout a sample, consistently with the Griffiths' theorem\cite{Griffiths_1968,Ruffo-Fanelli_book} for bulk magnets.  The compromise most often encountered in experiments is the occurrence of a multi-domain phase (highlighted by the ellipse in Fig.~\ref{mag_curve}).
The discontinuity marked in Fig.~\ref{mag_curve} with two bullets on the \textit{equilibrium} magnetization
curve produced by models with short-ranged interactions only is replaced by an analytic function  when dipolar interaction is taken into account.

A fundamental question then arises: along with the discontinuity in the equilibrium magnetization curve does dipolar interaction wipes away the critical behavior as well?
In the following we consider a minimal model and show that the dipolar interaction acts as a relevant field, in the meaning of the renormalization group, beside the reduced temperature and the external $B$ field\cite{Fisher_RevMod_1974,Stanley_RevMod_1999}.
This description is able to account for the coexistence of ordinary criticality with an analytic behavior of the magnetization as a function of $B$ at every temperature.  \\
In section II we introduce the model and summurize the main experimental facts that inspired our study. In section III we propose a scaling \textit{ansatz} that is validated in the forthcoming sections with a mean-field calculation (IV), a real-space renormalization group approach (V), and Monte-Carlo simulations (VI).

\section{Experimental facts and the model}
In this paper we address properties related to thermodynamic equilibrium with particular focus on the behavior of the magnetization as a function of the applied field $B$. Magnetic hysteresis, usually considered the distinctive feature of the magnetization curve of ferromagnets, is not an equilibrium phenomenon and is, therefore, beyond our scope.
Consistently with the scenario depicted in the introduction, the equilibrium magnetization for a real ferromagnet should behave smoothly as a function of the $B$ field when the latter passes from negative to positive values, without displaying any singularity at any temperature.
The experimental validation of this fact is usually precluded because magnets are normally not able to relax to the configuration of  minimal free energy within the measurement time. This means that magnetic hysteresis is also compatible with the vanishing of spontaneous magnetization at equilibrium
prescribed by the Griffith's theorem, as a result of dipolar frustration.
Together with other coworkers, we recently reported an experimental study in which the magnetization of a ferromagnet strongly frustrated by dipolar interaction was measured in a wide range of temperature ($T$) and applied $B$ field, taking care that hysteretic effects were negligible\cite{Saratz_Nature_2016}. This study was performed on Fe films epitaxially grown on Cu. For a thickness smaller than three atomic Fe layers these films are magnetized out of plane.  In this configuration the frustrating effect of dipolar interaction against ferromagnetism is maximal because the dipolar interaction between any pair of magnetic moments in the film is antiferromagnetic.
As a reference model for those Fe films we consider a 2D Ising Hamiltonian in which the usual nearest-neighbor ferromagnetic exchange interaction  competes with an antiferromagnetic interaction decaying with the third power of the distance. This second term arises from the isotropic contribution to pairwise dipolar coupling, the anisotropic contribution vanishing exactly in films magnetized out of plane. The model Hamiltonian thus reads
\begin{equation}
 H= - J \sum_{\langle i,j \rangle} S_i S_j + g\sum_{i \ne j} \frac{S_i
S_j}{r^3_{ij}} - h \sum_i S_i \label{Hamilton1}
\end{equation}
\noindent   where $S_i=\pm 1$ are Ising spins disposed on a square lattice, representing the two out of plane directions along which magnetic moments preferentially point. The first sum runs over every distinct pair of nearest-neighboring sites (positive exchange coupling constant $J>0$ is assumed henceforth). The second sum, associated with dipolar interaction with strength $g>0$, runs now over every distinct pair of sites in the lattice. The last term represents the Zeeman energy $h=\mu B$, with $\mu$ magnetic moment. As a result of the competition between the short- and long-ranged interaction,
this model exhibits modulated phases at low temperatures: striped phases at zero or small magnetic fields\cite{PiCa2007,ViSaPoPePo2008} and bubble phases at intermediate fields\cite{Portmann10PhysRevB,DiMu2010,Cannas_PRB_2011,Mendoza-Coto_PRL_2015}.
Due to this feature this model and its generalized versions, in which the antiferromangetic coupling decays with a generic exponent, have been studied extensively during the last two decades in relation to the type of order realized in the modulated phases (smectic, Ising nematic, etc.)
or to the possibility of producing self-generated glassiness\cite{Schmalian_2000_PRL,Nussinov_PRB_2004,Principi_PRL_2016}.
The uniform phase can be enforced by applying a magnetic field larger that a certain threshold value $h_{\rm c}$, which is generally temperature dependent\cite{Portmann10PhysRevB,Saratz_PRL2010}.
Whether some trace of the critical behavior, characterizing not frustrated models of ferromagnetism, is found in this uniform phase
has eluded scientific interest so far. We remark that in the frustrated model the global magnetization is not an order parameter in the conventional understanding of second-order phase transitions. It is therefore \textit{a priori} not obvious whether some critical behavior should be displayed at all in the uniform phase obtained when a large enough field is applied.
The experiments on Fe films on Cu confirmed without any doubt that ordinary criticality indeed occurs in the uniform phase of a strongly frustrated
ferromagnet. In that specific system the scaling behavior of the magnetization
\begin{equation}\label{scalingm1}
    m(\tau,B) = |\tau|^\beta\, F^{\pm}\left( \frac{B}{|\tau|^{\beta\delta}}\right)
\end{equation}
\noindent ($\tau=T/T_{\rm c}-1$)
is realized when external fields larger than a certain temperature-dependent threshold ($B_{\rm c}$) are applied and is consistent with the critical exponents $(\beta,\delta)$ and scaling functions  of the (unfrustrated) 2D Ising model.
The threshold field $B_{\rm c}$ varies from few Gauss for $\tau <-0.05$ to about 50 Gauss ($5\times 10^{-3}$ T) around $T_{\rm c}$).
In the -- so-called -- Griffiths-Widom representation power laws are obeyed up to eighty orders of magnitude\cite{Saratz_Nature_2016}.
The essence of the experimental observations on Fe films on Cu was confirmed by Monte-Carlo simulations  performed using the Hamiltonian~\eqref{Hamilton1},  even if realistic values of the Hamiltonian parameters could not be employed.
For instance, using a physical value for the ratio $J/g$ ($\sim 500$) would produce magnetic domains of size larger than the accessible simulation boxes. Therefore, the much smaller ratio  $J/g=10$ was used in order to observe both uniform and modulated phases in the same simulation.
In experiments, the fields that suffice to stabilize the uniform phase correspond to a Zeeman energy of the order of $10^{-5}$ times $J$,   
while for the parameters used in Monte-Carlo simulations these fields are about $10^{-1} - 10^{-2}$ times $J$ (see later on).
Differently from what done in the analysis of experimental results, the critical exponents of the 2D Ising model were assumed in the analysis of Monte-Carlo results
presented in Ref.~\onlinecite{Saratz_Nature_2016}.
Therein, the scaling behavior~\eqref{scalingm1} was verified adjusting the putative critical temperature $T_{\rm c}$ in order to obtain the maximal collapsing of simulated data.
The critical temperature deduced in this way showed a linear dependence on the strength of dipolar coupling $g$:
\begin{equation}
\label{T_C_MC}
 T_{{\rm c}, g} = T_{{\rm c}, 0}  - 11.3\, g
\end{equation}
where   $T_{{\rm c}, 0} = 2J/(\ln(1+\sqrt{2}))\approx 2.269\; J$ is the Onsager critical temperature.

These experimental and numerical facts suggest that the scaling hypothesis reported in textbooks of magnetism should be 
phrased in more general terms.

\section{The scaling hypothesis}

Both experimental and numerical evidence indicates that the critical behavior outside the multi-domain phase is controlled by the Onsager critical point, i.e. the critical point of the unfrustrated ferromagnet, ($g=0$) when $g/J \ll 1$.
This fact provides a valuable hint to set the scaling hypothesis~\eqref{scalingm1} in a broader framework.
Concretely, it suggests to replace the scaling function $F^{\pm}(x)$, which depends on a single variable, by a two-variable scaling function  $G^{\pm}(x,y)$.
This allows accounting for an additional  scaling field, parametrized by the variable
$u = g/J$ henceforth, that acts as a \textit{relevant} field for the unfrustrated critical point
$(T=T_{{\rm c}, 0},B=0)$,  in the sense of renormalization group.
Hence, we might assume that the magnetization $m(\tau,b,u)$, is a generalized homogeneous function satisfying the relation
\begin{equation}\label{scalingm2}
    m(\lambda^{1/\beta}\tau,\lambda^\delta b,\lambda^{1/\omega}u)= \lambda \, m(\tau,b,u)\;\;\; \;\;\; \forall\lambda \neq 0
\end{equation}
\noindent  where $b= h/J=\mu \,B/J$  and  $\omega$ is a new critical exponent. When $u=0$ we recover the behavior~\eqref{scalingm1} from Eq.~\eqref{scalingm2}. Choosing $\lambda ^{1/\omega}u=1$ we obtain the scaling form
\begin{equation}\label{scalingm3}
    m(\tau,b,u) = u^\omega G^{\pm}\left( \frac{|\tau|}{u^{\omega/\beta}},\frac{b}{u^{\omega\delta}}\right)
\end{equation}
\noindent where  $\pm$ refers to $\tau > 0$ or $\tau < 0$, respectively. The scaling functions $G^{\pm}(x,y)$ must satisfy some particular asymptotic behaviors.
For instance, when $u\neq0$ and $b\to 0$ the magnetization should vanish for any value of the temperature. From this follows the requirement
\begin{equation}
    \lim_{y\to 0} G^{\pm}(x,y) = 0.
\end{equation}
\noindent Moreover, Eq.~\eqref{scalingm1} must be recovered in the limit $u\to 0$ with $B$ finite, meaning that
\begin{equation}
     G^{\pm}(x,y) \sim x^{\beta} F^{\pm}\left( \frac{y}{x^{\beta\delta}}\right)
\end{equation}
\noindent for $x\gg 1$ and $y \gg 1$
with $y/x^{\beta\delta}$ finite. We then  expect two different scaling regimes depending on whether
\begin{equation}
\begin{split}
& b \ll b_{\rm c} \quad \Rightarrow \quad \text{multi-domain phase}\\
& b \gg b_{\rm c} \quad \Rightarrow \quad \text{ferromagnetic scaling region}
\end{split}
\end{equation}
the crossover field scaling with $u$ as $b_{\rm c} =h_{\rm c}/J \propto u^{\omega\delta}$.
In the first regime the equilibrium magnetization is a smooth function (essentially a straight line for Fe films on Cu) of the applied field. In the second regime ($ b \gg b_{\rm c}$)
standard criticality expressed by Eq.~\eqref{scalingm1} holds.
Defining $p(y)\equiv G^+(0,y)=G^-(0,y)$, for $\tau=0$ one has
\begin{equation}\label{scalingm4}
    \frac{m}{u^\omega}= p\left( \frac{b}{u^{\omega\delta}}\right) \,.
\end{equation}
Finally, to be compatible with Eq.~\eqref{scalingm2} the singular part of the free energy must also be a generalized homogeneous function and satisfy the relation
\begin{equation}\label{scalingf}
    f\left(\lambda^{\frac{1}{\beta(1+\delta)}}\tau,\lambda^{\frac{\delta}{1+\delta}} b,\lambda^{\frac{1}{\omega(1+\delta)}}u\right)= \lambda \, f(\tau,b,u)\;\;\; \;\;\; \forall\lambda \neq 0
\end{equation}

In the next sections  the validity of the  scaling \textit{ansatz} proposed in Eq.~\eqref{scalingm2} will be confirmed studying the model~\eqref{Hamilton1} by means of different
approaches: mean-field approximation, real-space renormalization group, Monte-Carlo simulations.

\section{Mean-field theory}

To analyze the mean-field approximation of Hamiltonian~\eqref{Hamilton1}, we consider the Landau-Ginzburg (LG) free energy

\begin{widetext}
\begin{equation}
\label{Hreal}
F[\phi] = \frac{1}{2} \int d^2 {\bf x } \left\{ \left( J\, \nabla
\phi({\bf x}) \right)^2 + r_0 \phi^2({\bf x}) + \frac{\bar \lambda}{2}
\phi^4({\bf x}) \right\} + \frac{g}{2} \int d^2 {\bf x} \int
d^2{\bf x}' \frac{\phi({\bf x})\phi({\bf x'})}{\left|{\bf x}-{\bf
x}'\right|^3} - h \int d^2 {\bf x
}\;  \phi({\bf x})
\end{equation}
\end{widetext}

\noindent where the scalar field $\phi({\bf x})$ represents the out-of-plane spin density
in a magnetic thin film with easy axis perpendicular to the film plane. Space variables ${\bf x}$ and ${\bf x'}$ are assumed to be dimensionless so that the constants
$J$, $r_0$, $\bar \lambda$, and $h$  have the units of an energy. The terms between curly brackets model the ferromagnetic exchange interactions
in the continuum limit. In the vicinity of the critical point one has
\begin{equation}\label{constLG}
\begin{split}
&r_0 = T-T_{{\rm c}, 0}^{\rm MF}    \\
&\bar \lambda = \frac{1}{3}T_{{\rm c},0}^{\rm MF}
\end{split}
\end{equation}
where $T_{{\rm c},0}^{\rm MF}$  is the mean-field transition temperature of the unfrustrated model  (with $g=0$), for which
the critical exponents are well-known:  $\beta=1/2$ and $\delta=3$.
Outside a limited region in the $(\tau,h)$ parameters space, i.e. for $h>h_{\rm c}$, domain states are not stable (they are either metastable or unstable) and the uniform solution $\phi=\phi_0$ is the equilibrium one\cite{Cannas_PRB_2011}. The saddle-point equation for $\phi_0$ is
\begin{equation}\label{saddle1}
(r_0+a_g\, g)\, \phi_0 + \phi_0^3 = h
\end{equation}
\noindent where
\begin{equation}
\label{ag_const}
a_g=\int \frac{ d^2 {\bf x}  }{\left|{\bf x}\right|^3} \,.
\end{equation}
\noindent One can implicitly assume a lower cutoff so that the integral above is not ill-defined. However, in the original functional~\eqref{Hreal}
the dipolar interaction is well-defined in the domain of distributions. An explicit value can be assigned to the constant $a_g$
considering the coarse-grained (magnetostatic) description of the equivalent model, that is
a slab of volume $V$ uniformly magnetized out of plane. For this system the demagnetizing energy is
\begin{equation}
\label{demagEnergy}
\mathcal{E}_{\rm d}=\frac{1}{2}\mu_0 M^2 V=\frac{1}{2}\mu_0 \mu^2 \left(\frac{\phi_0}{a^3}\right)^2 V \,.
\end{equation}
In the last equivalence a cubic lattice of constant $a$ has been assumed as well as the obvious relation
$M=\mu\, \phi_0/a^3$  between the macroscopic magnetization $M$ and the uniform spin density $\phi_0$ considered here.
The demagnetizing energy in Eq.~\eqref{demagEnergy} should equal the corresponding contribution in the LG functional
\begin{equation}
\begin{split}
 \label{H_d}
F_{\rm d}[\phi]& = \frac{g}{2}
\int d^2 {\bf x} \int
d^2{\bf x}' \frac{\phi({\bf x})\phi({\bf x'})}{\left|{\bf x}-{\bf
x}'\right|^3} \\
&=\frac{1}{2} \, \frac{\mu_0}{4\pi} \,  \frac{\mu^2}{a^3}  \,  \phi_0^2
\int d^2{\bf x}'\int \frac{  d^2 {\bf x}  }{\left|{\bf x}\right|^3}
\end{split}
\end{equation}
where in the second passage we have expressed the coupling constant $g$ in terms of the atomic magnetic moment and the lattice unit.
A one-to-one mapping can now be established between individual terms in the Eqs.~\eqref{demagEnergy}  and ~\eqref{H_d}
\begin{equation}
\begin{split}
& \frac{\mu}{a^3}\,  \phi_0  = M\\
& \mu \,\phi_0 \int d^2 {\bf x}' = MV \\
& \frac{1}{4\pi} \int  \frac{ d^2 {\bf x}  }{\left|{\bf x}\right|^3} =1 \,.
\end{split}
\end{equation}
Therefore, in this description, the constant in Eq.~\eqref{ag_const} is $a_g=4\pi$.
From the saddle-point Eq.~\eqref{saddle1} associated with a uniform spin density $\phi_0$, we note that
the presence of dipolar interaction effectively lowers the critical temperature by an amount  $4\pi\,g$, namely
\begin{equation}
\label{T_C_MF}
 T_{{\rm c},g}^{\rm MF} = T_{{\rm c}, 0}^{\rm MF}  - 4\pi \, g\,.
\end{equation}
Even if deduced in a mean-field context, the correction to the critical temperature provided by dipolar coupling is  in excellent agreement with the results of Monte-Carlo simulations on a square lattice summarized in Eq.~\eqref{T_C_MC}.

Let us go back to the main purpose of this section of verifying the validity of our scaling hypothesis within the mean-field approximation.
To this aim, we define $z=\phi_0/u^{1/2}$ and divide both sides of the saddle-point Eq.~\eqref{saddle1} by $u^{3/2}$ to obtain
\begin{equation}\label{saddle2}
    \left(a_0\frac{\tau}{u}+a_1\right)\, z + z^3 = \frac{h}{u^{3/2}}
\end{equation}
where $a_0$ and $a_1$ are numerical constants.
\noindent Therefore,  we obtain
\begin{equation}
\phi_0 = u^{1/2} G^{\pm}\left( \frac{|\tau|}{u},\frac{h}{u^{3/2}}\right)
\end{equation}

\noindent which is consistent with the scaling hypothesis proposed in~\eqref{scalingm3}, with a mean-field dipolar critical exponent $\omega=1/2$
characterizing the \textit{dipolar} relevant field $u$.

\section{Real-space renormalization group approach (RSRG)}
\begin{figure}
\begin{center}
\includegraphics[width=9cm]{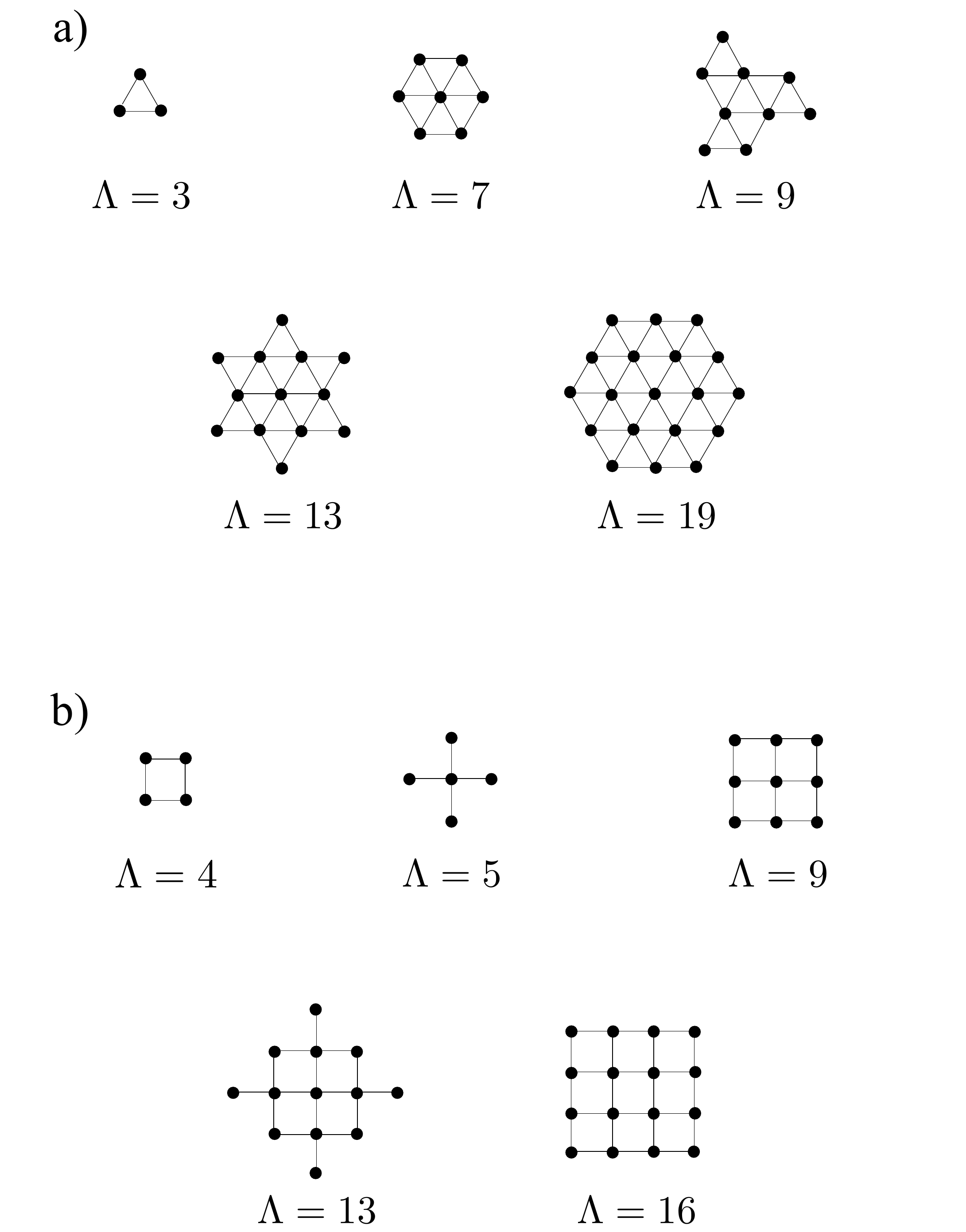}
\caption{\label{blocks} Kadanoff blocks for different values of $\Lambda$ used in the RG approach for (a) triangular lattice and (b) square lattice.}
\end{center}
\end{figure}

An fundamental requirement for the consistency of the scaling hypothesis in Eq.~\eqref{scalingm2} is that the variable $u$  be a relevant field within a renormalization-group (RG) approach.  In the following we demonstrate that this is indeed the case using the Niejmeijer and van Leeuwen RSRG technique\cite{NvL} and its extension to include long-range interactions\cite{Cannas1995}. For the sake of simplicity, we prove this only for the $B=0$ case, the extension to $B\neq 0$ being straightforward. Defining ${\cal H} \equiv -\beta H$, from Eq.~\eqref{Hamilton1}  we have
\begin{equation}
{\cal H} = K_1 \sum_{\langle i,j \rangle} S_i S_j - K_2 \sum_{i \ne j} \frac{S_i
S_j}{r^3_{ij}}  \label{HamiltonRG}
\end{equation}

\noindent namely, $K_1=\beta J$ and $K_2=\beta g$. We divide the system into Kadanoff Blocks with $\Lambda$ spins, so that the rescaling length of the RG transformation is $l=\sqrt{\Lambda}$. To each block $I$ we assign a block spin $S'_I=\pm 1$. Defining the renormalized block Hamiltonian as

 \begin{equation}
{\cal H}'= K'_1 \sum_{\langle  I,J \rangle}  S'_I S'_J - K'_2 \sum_{I \ne J} \frac{S'_I
S'_J}{r^3_{IJ}}  \label{HamiltonRG2}
\end{equation}

\noindent where  $r_{IJ}$ is the distance between blocks $I$ and $J$ measured in units of the rescaled length $l$, we obtained the recursion RG equations
\begin{eqnarray}
\label{RG_recursion}
  K'_1 &=& K'_1(K_1,K_2) \nonumber  \\
  K'_2 &=& K'_1(K_1,K_2)
\end{eqnarray}

\noindent for each block set. In Fig.~\ref{blocks} we show the different Kadanoff blocks used in the RG calculation.  The details of the RG implementation are given in Appendix~\ref{Ap1}.
It is immediate to see that $K_2=0$ implies $K'_2=0$. Therefore,
the non-trivial fixed point of the RG equations~\eqref{RG_recursion} is located at  $(K_{\rm c},0)$, with $K_{\rm c}$ determined by the equation $K'_1(K_{\rm c},0)=K_{\rm c}$.  
In the RG approach, this fixed point corresponds to the critical temperature, i.e., $T_{\rm c,0}^{\rm RG}=J/k_B K_{\rm c}$ is the transition temperature of the unfrustrated model under the present approximation.
Still in zero magnetic field, the RG equations linearized around this critical point are given by the matrix
\begin{equation*}
\left(\begin{array}{cc}
        \frac{\partial K'_1}{\partial K_1} & \frac{\partial K'_1}{\partial K_2} \\
        \frac{\partial K'_2}{\partial K_1} & \frac{\partial K'_2}{\partial K_2}
      \end{array}
 \right)_{K_1=K_{\rm c},K_2=0}=
 \left(\begin{array}{cc}
        \lambda_\tau & \frac{\partial K'_1}{\partial K_2} \\
        \frac{\partial K'_2}{\partial K_1} & \frac{\partial K'_2}{\partial K_2}
      \end{array}
 \right)_{K_1=K_{\rm c},K_2=0}
\end{equation*}
where $\lambda_\tau$  is the so-called thermal eigenvalue (see Appendix~\ref{Ap1}).
From Eq.~\eqref{Kp2} we have
\begin{equation*}
\left. \frac{\partial K'_2}{\partial K_1}\right|_{K_1=K_{\rm c},K_2=0}=0
\end{equation*}
\noindent and, therefore, the eigenvalue associated with the dipolar-coupling constant (in units of $k_{\rm B}T$)  $K_2$ is given by
\begin{equation}\label{lambdau}
\lambda_u= \left. \frac{\partial K'_2}{\partial K_2}\right|_{K_1=K_{\rm c},K_2=0}=  \frac{1}{l^3} \, \left[ M(K_{\rm c}) \right]^2
\end{equation}
\noindent where $M(K)$ is given in Eq.~\eqref{MK}. It is also easy to see that the field eigenvalue is given by $\lambda_b=M(K_{\rm c})$.

The main idea behind the present approach is that eigenvalues of the RG equations linearized around a non-trivial fixed point that are larger than one
correspond to \textit{relevant fields}\cite{LeBellac,Fisher_RevMod_1974,Stanley_RevMod_1999}. The aim of this section is, thus, to prove that $\lambda_u>1$. 
Typically, eigenvalues associated with relevant fields display a power-law dependence on the size of the Kadanoff blocks that are specific of the implemented  renormalization procedure\cite{Delamotte_Springer_2012}.
The relative exponents are directly related to physical critical exponents (see below).

We calculated the eigenvalues for different values of $\Lambda$, both for the triangular and the square lattices shown in Fig.~\ref{blocks}.  Since the critical exponents are expected to be independent of the lattice structure, we could expect the general trend of the eigenvalues with $l$ to be the same for both lattices.
We now make a change of variables from $(K_1,K_2)$ to $(\tau,u)$, with $u=K_2/K_1$, $\tau=K_{\rm c}/K_1-1$. Then, neglecting in first approximation the non-diagonal element of the RG matrix   and assuming
\begin{eqnarray*}
  \lambda_\tau&\sim& l^{y_\tau}\\
  \lambda_b&\sim& l^{y_b}\\
  \lambda_u&\sim& l^{y_u}
\end{eqnarray*}
(where we have included now a finite magnetic field) it is easy to show that the singular part of the free energy should scale, close to the critical point, as
\begin{equation}
f(l^{y_\tau/d}\tau,l^{y_b/d} b, l^{y_u/d} u) \approx l\; f(\tau,b,u)\,,
\end{equation}
\noindent with $d=2$ being the dimensionality of the system. Hence, comparing the equation above with Eq.~\eqref{scalingf} one obtains that
\begin{eqnarray}
  y_\tau &=& \frac{2}{\beta(1+\delta)}=\frac{1}{\nu} \label{yT}\\
  y_b &=& \frac{2\delta}{1+\delta} \label{yB}\\
  y_u &\approx& \frac{2}{\omega(1+\delta)} \label{yu} \,.
\end{eqnarray}
\noindent While the first two equations are  well-known, to the best of our knowledge, the third Eq.~\eqref{yu} is not reported in the literature.  This last equation
allows relating the dipolar critical exponent $\omega$ to the other exponents.

\begin{figure}
\begin{center}
\includegraphics[scale=0.42]{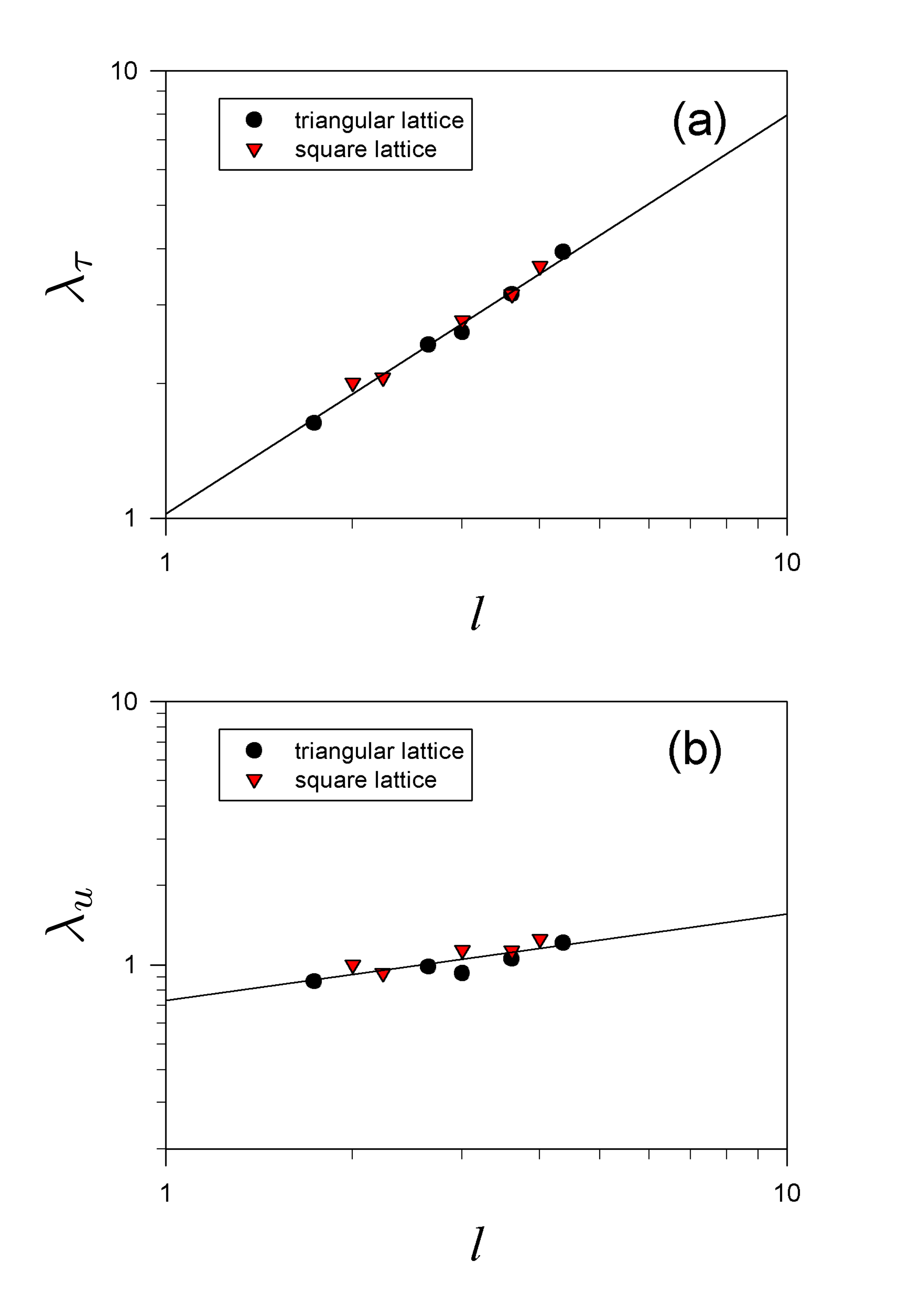}
\caption{\label{lambdaT} Thermal and dipolar eigenvalues  vs. the scale length $l$. (a) $\lambda_\tau$. (b) $\lambda_u$}
\end{center}
\end{figure}

As a consistency check of our RG approach, we first calculated $\lambda_\tau$ as a function of $l=\sqrt{\Lambda}$. Fig.~\ref{lambdaT}a shows that $\lambda_\tau$ displays the expected power-law behavior (line in a log-log scale). A linear fitting combining the data for both the square and the triangular lattice yields the exponent $y_\tau=0.89$, reasonably close to the exact value $y_\tau=1$.
We then proceeded considering the behavior of the eigenvalue associated with the dipolar coupling, of our interest.
Fig.~\ref{lambdaT}b shows $\lambda_u$ as a function of $l$. As one can see, $\lambda_u$ is actually smaller than one for small values of $l$,  because  $\lambda_u =\lambda_0\,l^{y_u}$  with $\lambda_0<1$. However, also this eigenvalue clearly obeys a power-law behavior, which indicates that $\lambda_u$ becomes larger than one
when sufficiently large Kadanoff blocks are considered\cite{Cannas1995}.
In this sense, the crucial result of this section is that $y_u>0$, which confirms the assumption of $u$ being a relevant field. The exponent resulting from the fit of the $\lambda_u$
eigenvalues computed for different lattices is $y_u=0.33$. From Eqs.(\ref{yT})-(\ref{yu}) this implies  $\omega=1.015$.
For completeness, it is worth mentioning that the same calculation for the field eigenvalue $\lambda_b$ yields a critical exponent $\delta=4.97$, very different from the exact result $\delta=15$. Hence, the value of $\omega$ resulting from this RG calculation can deviate significantly from
the estimate of the same critical exponent obtained from Monte-Carlo simulations.

Finally, we note that the non--diagonal structure of the RG matrix  implies that the eigenvectors associated with the eigenvalues $ \lambda_\tau$ and 
$ \lambda_u$ are actually not orthogonal. Taking this into account\cite{NvL},  would produce a correction in 
the critical temperature $T_{\rm c,0}^{\rm RG}=J/k_B K_{\rm c}$ which  scales linearly  with the strength of dipolar coupling $g$. Even without developing the calculation in details, the outcome would then be consistent with our Monte-Carlo and mean-field results (see Eqs.~(\ref{T_C_MC}, \ref{T_C_MF})). 
A second consequence of the non--orthogonality of eigenvectors is a correction in  Eq.~\eqref{yu}, which could lead to an improved estimate of the 
critical exponent $\omega$.

\section{Monte-Carlo simulations}

Monte-Carlo (MC) simulations were performed using Hamiltonian ~\eqref{Hamilton1} on a square lattice comprising $N=L\times L$ sites, with  $L=200$ for all simulations; periodic boundary conditions were assumed and handled implementing Ewald sums. We calculated the total magnetization $m=\sum \langle S_i\rangle/N$ (where $\langle S_i\rangle$ stands for the statistical average) as a function of $h$ at $T=T_{{\rm c}, 0} \approx 2.27 J$ for $h>h_{\rm c}$ and different different values of $u=g/J$. The critical field $h_{\rm c}$  was estimated, either by direct calculation of the corresponding order parameters (for small values of $J/g$) or by a zero-field cooling -- field-cooling procedure, as described in Ref.~\onlinecite{Saratz_Nature_2016}. At the beginning of each MC run  we let the system equilibrate over $1000$ Monte-Carlo Steps (MCS) and then average over $1000$ sampling point taken every $100$ MCS along a single MC run.

\begin{figure}[h]
\begin{center}
\includegraphics[scale=0.4]{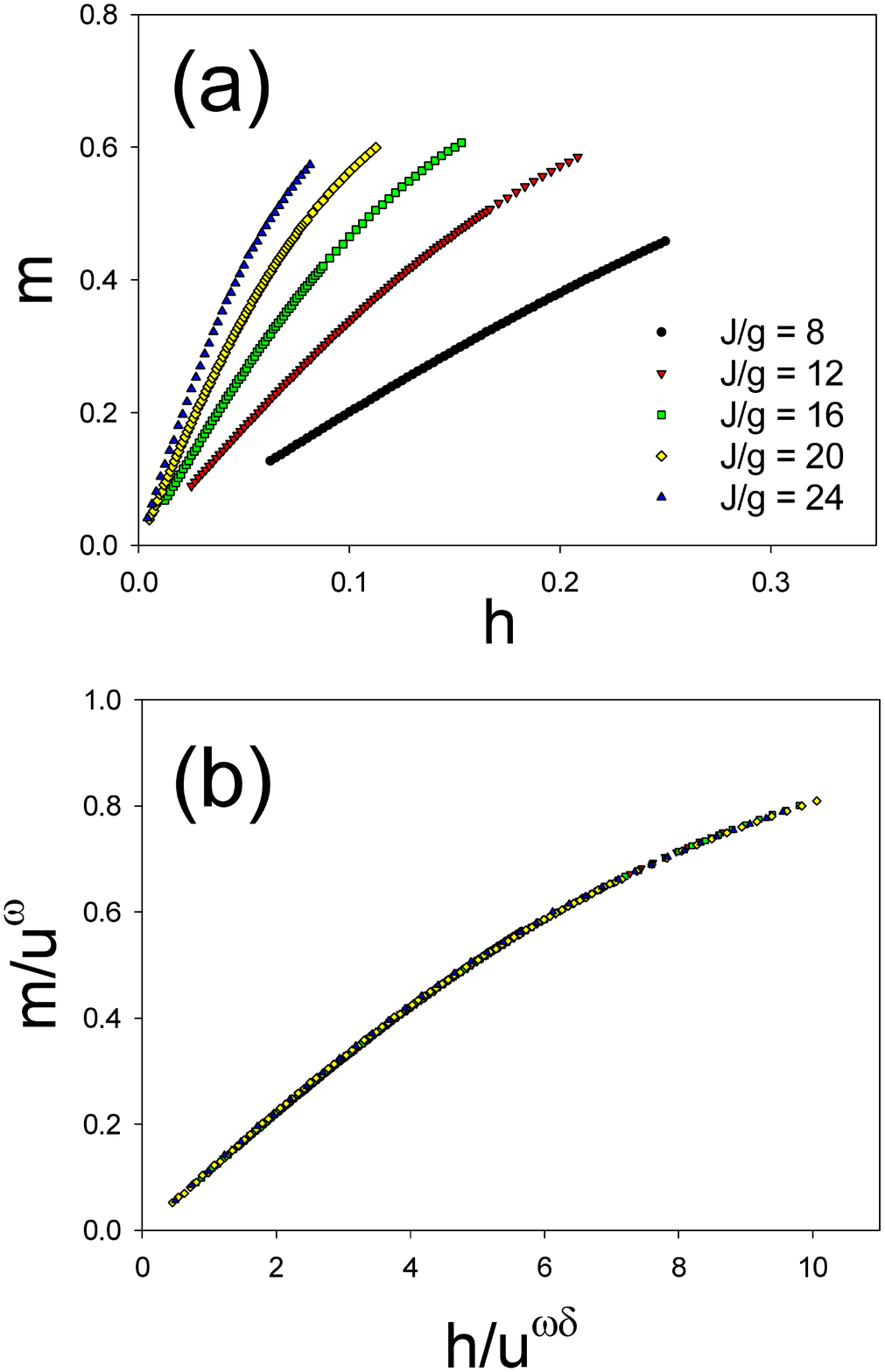}
\caption{\label{scalingu} Monte-Carlo simulations for $L=200$, $T=T_{{\rm c}, 0}$ and different values of $J/g$. (a) Magnetization as a function of $h$. Data collapse of the same magnetization curves for $\omega = 0.1 \pm 0.002$ ($\delta=15$).}
\end{center}
\end{figure}

In Fig.\ref{scalingu}a we show the simulated magnetization curves as a function of $h$ for different values of $J/g$ computed at $T=T_{{\rm c}, 0}$.
In Fig.\ref{scalingu}b we show a scaling plot of $m/u^\omega$ vs. $h/u^{\omega\delta}$ for the same data set assuming $\omega\approx 1/10$ and $\delta=15$. The excellent collapsing of data confirms the scaling relation~\eqref{scalingm4}, which descends directly from the proposed scaling hypothesis~\eqref{scalingm3}.

\section{Discussion}

We proposed an extension of the textbook scaling  \textit{ansatz}  for ferromagnets that applies
to the realistic situation in which the formation of magnetic domains -- promoted by dipolar interaction -- renders the Curie point technically unreachable.
This \textit{ansatz} is based on the assumption that the dipolar coupling acts as a relevant field, in the sense of renormalization group. This implies that a dipolar critical exponent $\omega$ needs to be introduced, besides the traditional
ones related to the ferromagnetic-to-paramagnetic phase transition.
The most reliable estimate of this exponent is the one resulting from Monte-Carlo simulations, i.e., $\omega \approx 1/10$.
In fact, in this respect, the accuracy of mean-field theory and real-space renormalization-group approach is notoriously poor even for the unfrustrated model\cite{LeBellac,Delamotte_Springer_2012}.
However, both these analytic approaches support the basic assumption of dipolar coupling being a relevant field.

Since dipolar interactions are ubiquitous and unavoidable in real magnets, our results suggest that long-range ferromagnetic order should be regarded as a crossover phenomenon. In other words, in the realm of equilibrium thermodynamics, the scaling behavior associated with the onset of long-range ferromagnetic ordering should be observable in the neighborhood of the putative critical point, but not too close to it. In fact, when the latter is approached by letting all the relevant fields $(\tau,b,u)$ go to zero, phases with modulated magnetization intervene that display a non-singular behavior of the ferromagnetic order parameter ($m$).   
In this perspective, our results reconcile the Griffith's theorem\cite{Griffiths_1968,Ruffo-Fanelli_book} and dipolar frustration with the observation of criticality, at least for ferromagnetic films magnetized out of plane.

We hope that the present work will stimulate further investigations aimed at validating the proposed scaling hypothesis~\eqref{scalingm3} for 3D magnets.  
Moreover, the model studied by us belongs to a more general class of models in which a generic exponent  $\alpha$  is assumed for the power decay of the long-range interaction\cite{Giuliani_PRB_11,Giuliani_PRB_07,Mendoza-Coto_PRL_2015,Viot}.      
The theoretical scenario of self-generated phase separation
\footnote{In applications of these toy models to high-$T_{\rm c} $ superconductors 
it is assumed that the short-range attraction among holes arises from the presence of an underlying antiferromagnetic phase. The competition 
between this attractive interaction and Coulomb repulsion then produces hole-rich regions alternating with hole-poor antiferromagnetic regions within a sample, equivalent to magnetic domains in dipolar-frustrated systems\cite{Keimer_Nature_2015,Emery_PhysC_1993}.}  
and avoided criticality  was originally proposed for one of these models (with $\alpha=1$) in the context of high-$T_{\rm c} $ superconductors\cite{Tranquada_Nature_1995,Emery_PhysC_1993}.  
The approach presented here could potentially help understand the complex phase diagram of this second class of materials\cite{Keimer_Nature_2015}.

\acknowledgments

This work was partially supported by  CONICET (Argentina) through grant
PIP 11220150100285 and  SeCyT (Universidad Nacional de C\'ordoba, Argentina). We thank fruitful suggestions from S. Bustingorry and T. Grigera. Danilo Pescia is acknowledged for putting this interesting open problem under our attention and for several illuminating discussions we had with him on this topic.

\appendix

\section{Real-space renormalization group implementation \label{Ap1}}

Following Niejmeijer and van  Leeuwen prescription, we first divided the Hamiltonian~\eqref{HamiltonRG} into two parts: ${\cal H}= {\cal H}_0+{\cal V}$, where ${\cal H}_0=\sum_I {\cal H}_0^I$ and ${\cal V}= \sum_{ I\ne J} {\cal V}_{IJ}$; ${\cal H}_0^I$ includes only the interactions between spins inside the block $I$, whereas ${\cal V}_{IJ}$ includes the interactions between spins belonging to different blocks $I$ and $J$. We also denoted $S_i^I$
($i=1,\ldots, \Lambda$) the site spins belonging to the block $I$.  The renormalized Hamiltonian~\eqref{HamiltonRG2}, in the first order cumulant approximation\cite{NvL,Cannas1995} is then given by
\begin{equation}\label{Hp}
    {\cal H}'= \sum_{I \ne J} \langle {\cal V}_{IJ} \rangle_0,
\end{equation}

\noindent where

\begin{equation}
\label{average}
\langle {\cal O} \rangle_0 =\frac{1}{Z_0}\; {\rm Tr}_{\{S_i^I\}} P\left(\{S_i^I\},\{ S'_I\}\right)\, \exp \left[ {\cal H}_0(\{ S_i^I \}) \right]\, {\cal O}
\end{equation}

\noindent with

\[
Z_0= {\rm Tr}_{\{S_i^I\}} P\left(\{S_i^I\},\{S'_I\}\right)\, \exp \left[{\cal H}_0(\{S_i^I\})\right]
\]

\noindent and

\[
P\left(\{S_i^I\},\{S'_I\}\right) = \prod_I \, \frac{1}{2} \left[1+ S'_I\, {\rm sgn} \left(\sum_{i=1}^\Lambda S_i^I \right) \right]
\]

\noindent is the weight function which characterizes the majority rule recipes. We will use this expression also when $ \Lambda$ is an even number, meaning that $P$ assigns the values  $S'_I=\pm 1$ with probability $1/2$ to  spins configurations with zero magnetization in the block. In particular, it is easy to see that $\langle S_i^I \rangle_0= a_i(\vec{K})\, S'_I$, where $a_i(\vec{K})$ does not depend on the block $I$. Assuming now that\cite{Cannas1995} $r_{ij} \approx l\, r_{IJ}$ for $r_{IJ}>1$, replacing into Eqs.(\ref{HamiltonRG}) and (\ref{Hp}), and comparing with Eq.(\ref{HamiltonRG2}), after some straightforward algebra  we find

\begin{equation}
  K'_2 = \frac{K_2}{l^3} \, \left[ \sum_{i \in I} a_i(\vec{K}) \right]^2\label{Kp2}
\end{equation}

\begin{equation}
  K'_1 = K_1 \,  \sideset{}{'}\sum_{i \in I}  \sideset{}{'}\sum_{j \in J}  a_i a_j -K_2  \sum_{i \in I}\sum_{j \in J} \frac{1}{r_{ij}^3} a_i a_j + K'_2\label{Kp1}
\end{equation}

\noindent where $I$ and $J$ in the last equation are nearest-neighboring blocks. The first pair of sums (primed sums) in Eq.(\ref{Kp1}) run over nearest-neighboring sites $i$ and $j$, while the second pair run over all sites in  both blocks.  Other useful block dependent quantities were calculated, such as

\begin{equation}
L(K)\equiv \left.  \sideset{}{'}\sum_{i \in I}  \sideset{}{'}\sum_{j \in J}  a_i a_j\right|_{K_1=K,K_2=0}
\end{equation}

\noindent and

\begin{equation}
M(K)\equiv \left. \sum_{i \in I}   a_i(\vec{K})\right|_{K_1=K,K_2=0}
\label{MK}
\end{equation}

\noindent For instance, the ferromagnetic (short range) critical point is determined by $L(K_{\rm c})=1$ and the corresponding thermal  eigenvalue $\lambda_\tau=l^{1/\nu}$ by

\begin{equation}
    \lambda_\tau = \left. \frac{\partial K'_1}{\partial K_1}\right|_{K_1=K_{\rm c},K_2=0}= 1+K_{\rm c} \left. \frac{dL}{dK}\right|_{K=K_{\rm c}}
\end{equation}

We see that, to determine the stability of the ferromagnetic fixed point under the present approximation we just need the quantities 

\begin{equation}\label{aa}
a'_i(K)=a_i(K,0)=\left.\langle S_i^I \rangle_0\right|_{S'_I=1,K_1=K,K_2=0}.
\end{equation}

 Let's consider a simple example for $\Lambda=5$, which corresponds to a cross-shaped Kadanoff block (see Fig.\ref{blocks}b). Suppose that we label $i=0$ the central site and $i=1,2,3,4$ the external sites of the block. By symmetry,  the coefficients $a'_i(K)=a_e(K)$, $i=1,2,3,4$,  are all equivalent. Then from Eqs.(\ref{average}) and (\ref{aa}) we obtain
\begin{equation}
a_e(K)=\frac{e^{4K}+e^{-4K}+2e^{2K}+2e^{-2K}}{6+e^{4K}+e^{-4K}+4e^{2K}+4e^{-2K}}
\end{equation}
\noindent 
and for the central site
\begin{equation}
a'_0(K)=\frac{6+e^{4K}-e^{-4K}+4e^{2K}-4e^{-2K}}{6+e^{4K}+e^{-4K}+4e^{2K}+4e^{-2K}} \,.
\end{equation}

Between two neighboring blocks there are three first-neighbor bonds. Hence, $L(K)=3[a_e(K)]^2$ and $M(K)=4a_e(K)+a_0(K)$. For large clusters the
 functions $a_i(K)$ can be obtained with the aid of symbolic manipulation programs, for clusters of size up to $\Lambda \approx 20$ (see Fig.\ref{blocks}).

\bibliographystyle{apsrev}


\begin{thebibliography}{27}
\expandafter\ifx\csname natexlab\endcsname\relax\def\natexlab#1{#1}\fi
\expandafter\ifx\csname bibnamefont\endcsname\relax
  \def\bibnamefont#1{#1}\fi
\expandafter\ifx\csname bibfnamefont\endcsname\relax
  \def\bibfnamefont#1{#1}\fi
\expandafter\ifx\csname citenamefont\endcsname\relax
  \def\citenamefont#1{#1}\fi
\expandafter\ifx\csname url\endcsname\relax
  \def\url#1{\texttt{#1}}\fi
\expandafter\ifx\csname urlprefix\endcsname\relax\def\urlprefix{URL }\fi
\providecommand{\bibinfo}[2]{#2}
\providecommand{\eprint}[2][]{\url{#2}}

\bibitem[{\citenamefont{Taroni et~al.}(2008)\citenamefont{Taroni, Bramwell, and
  Holdsworth}}]{Taroni_2008}
\bibinfo{author}{\bibfnamefont{A.}~\bibnamefont{Taroni}},
  \bibinfo{author}{\bibfnamefont{S.~T.} \bibnamefont{Bramwell}},
  \bibnamefont{and} \bibinfo{author}{\bibfnamefont{P.~C.~W.}
  \bibnamefont{Holdsworth}}, \bibinfo{journal}{Journal of Physics: Condensed
  Matter} \textbf{\bibinfo{volume}{20}}, \bibinfo{pages}{275233}
  (\bibinfo{year}{2008}).

\bibitem[{\citenamefont{Stanley}(1999)}]{Stanley_RevMod_1999}
\bibinfo{author}{\bibfnamefont{H.~E.} \bibnamefont{Stanley}},
  \bibinfo{journal}{Rev. Mod. Phys.} \textbf{\bibinfo{volume}{71}},
  \bibinfo{pages}{S358} (\bibinfo{year}{1999}).

\bibitem[{\citenamefont{Goldenfeld}(1992)}]{Goldenfeld}
\bibinfo{author}{\bibfnamefont{N.}~\bibnamefont{Goldenfeld}},
  \emph{\bibinfo{title}{Lectures on phase transitions and the renormalization
  group}}, vol.~\bibinfo{volume}{85} of \emph{\bibinfo{series}{Frontiers in
  physics}} (\bibinfo{publisher}{Perseus Books Publishing},
  \bibinfo{address}{Reading, Massachusetts}, \bibinfo{year}{1992}).

\bibitem[{\citenamefont{Le~Bellac}(1991)}]{LeBellac}
\bibinfo{author}{\bibfnamefont{M.}~\bibnamefont{Le~Bellac}},
  \emph{\bibinfo{title}{Quantum and Statistical Field Theory}}, Oxford Science
  Publ (\bibinfo{publisher}{Clarendon Press}, \bibinfo{address}{Oxford},
  \bibinfo{year}{1991}).

\bibitem[{\citenamefont{Griffiths}(1968)}]{Griffiths_1968}
\bibinfo{author}{\bibfnamefont{R.~B.} \bibnamefont{Griffiths}},
  \bibinfo{journal}{Phys. Rev.} \textbf{\bibinfo{volume}{176}},
  \bibinfo{pages}{655} (\bibinfo{year}{1968}).

\bibitem[{\citenamefont{Campa et~al.}(2014)\citenamefont{Campa, Dauxois,
  Fanelli, and Ruffo}}]{Ruffo-Fanelli_book}
\bibinfo{author}{\bibfnamefont{A.}~\bibnamefont{Campa}},
  \bibinfo{author}{\bibfnamefont{T.}~\bibnamefont{Dauxois}},
  \bibinfo{author}{\bibfnamefont{D.}~\bibnamefont{Fanelli}}, \bibnamefont{and}
  \bibinfo{author}{\bibfnamefont{S.}~\bibnamefont{Ruffo}},
  \emph{\bibinfo{title}{Physics of Long-Range Interacting Systems}}
  (\bibinfo{publisher}{Oxford University Press}, \bibinfo{address}{Oxford},
  \bibinfo{year}{2014}).

\bibitem[{\citenamefont{Fisher}(1974)}]{Fisher_RevMod_1974}
\bibinfo{author}{\bibfnamefont{M.~E.} \bibnamefont{Fisher}},
  \bibinfo{journal}{Rev. Mod. Phys.} \textbf{\bibinfo{volume}{46}},
  \bibinfo{pages}{597} (\bibinfo{year}{1974}).

\bibitem[{\citenamefont{Saratz et~al.}(2016)\citenamefont{Saratz, Zanin,
  Ramsperger, Cannas, Pescia, and Vindigni}}]{Saratz_Nature_2016}
\bibinfo{author}{\bibfnamefont{N.}~\bibnamefont{Saratz}},
  \bibinfo{author}{\bibfnamefont{D.~A.} \bibnamefont{Zanin}},
  \bibinfo{author}{\bibfnamefont{U.}~\bibnamefont{Ramsperger}},
  \bibinfo{author}{\bibfnamefont{S.~A.} \bibnamefont{Cannas}},
  \bibinfo{author}{\bibfnamefont{D.}~\bibnamefont{Pescia}}, \bibnamefont{and}
  \bibinfo{author}{\bibfnamefont{A.}~\bibnamefont{Vindigni}},
  \bibinfo{journal}{Nat. Commun.} \textbf{\bibinfo{volume}{7}},
  \bibinfo{pages}{13611} (\bibinfo{year}{2016}).

\bibitem[{\citenamefont{Pighin and Cannas}(2007)}]{PiCa2007}
\bibinfo{author}{\bibfnamefont{S.~A.} \bibnamefont{Pighin}} \bibnamefont{and}
  \bibinfo{author}{\bibfnamefont{S.~A.} \bibnamefont{Cannas}},
  \bibinfo{journal}{Phys. Rev. B} \textbf{\bibinfo{volume}{75}},
  \bibinfo{pages}{224433} (\bibinfo{year}{2007}).

\bibitem[{\citenamefont{Vindigni et~al.}(2008)\citenamefont{Vindigni, Saratz,
  Portmann, Pescia, and Politi}}]{ViSaPoPePo2008}
\bibinfo{author}{\bibfnamefont{A.}~\bibnamefont{Vindigni}},
  \bibinfo{author}{\bibfnamefont{N.}~\bibnamefont{Saratz}},
  \bibinfo{author}{\bibfnamefont{O.}~\bibnamefont{Portmann}},
  \bibinfo{author}{\bibfnamefont{D.}~\bibnamefont{Pescia}}, \bibnamefont{and}
  \bibinfo{author}{\bibfnamefont{P.}~\bibnamefont{Politi}},
  \bibinfo{journal}{Phys. Rev. B} \textbf{\bibinfo{volume}{77}},
  \bibinfo{pages}{092414} (\bibinfo{year}{2008}).

\bibitem[{\citenamefont{Portmann et~al.}(2010)\citenamefont{Portmann, G\"olzer,
  Saratz, Billoni, Pescia, and Vindigni}}]{Portmann10PhysRevB}
\bibinfo{author}{\bibfnamefont{O.}~\bibnamefont{Portmann}},
  \bibinfo{author}{\bibfnamefont{A.}~\bibnamefont{G\"olzer}},
  \bibinfo{author}{\bibfnamefont{N.}~\bibnamefont{Saratz}},
  \bibinfo{author}{\bibfnamefont{O.~V.} \bibnamefont{Billoni}},
  \bibinfo{author}{\bibfnamefont{D.}~\bibnamefont{Pescia}}, \bibnamefont{and}
  \bibinfo{author}{\bibfnamefont{A.}~\bibnamefont{Vindigni}},
  \bibinfo{journal}{Phys. Rev. B} \textbf{\bibinfo{volume}{82}},
  \bibinfo{pages}{184409} (\bibinfo{year}{2010}).

\bibitem[{\citenamefont{Diaz-Mendez and Mulet}(2010)}]{DiMu2010}
\bibinfo{author}{\bibfnamefont{R.}~\bibnamefont{Diaz-Mendez}} \bibnamefont{and}
  \bibinfo{author}{\bibfnamefont{R.}~\bibnamefont{Mulet}},
  \bibinfo{journal}{Phys. Rev. B} \textbf{\bibinfo{volume}{81}},
  \bibinfo{pages}{184420} (\bibinfo{year}{2010}).

\bibitem[{\citenamefont{Cannas et~al.}(2011)\citenamefont{Cannas, Carubelli,
  Billoni, and Stariolo}}]{Cannas_PRB_2011}
\bibinfo{author}{\bibfnamefont{S.~A.} \bibnamefont{Cannas}},
  \bibinfo{author}{\bibfnamefont{M.}~\bibnamefont{Carubelli}},
  \bibinfo{author}{\bibfnamefont{O.~V.} \bibnamefont{Billoni}},
  \bibnamefont{and} \bibinfo{author}{\bibfnamefont{D.~A.}
  \bibnamefont{Stariolo}}, \bibinfo{journal}{Phys. Rev. B}
  \textbf{\bibinfo{volume}{84}}, \bibinfo{pages}{014404}
  (\bibinfo{year}{2011}).

\bibitem[{\citenamefont{Mendoza-Coto et~al.}(2015)\citenamefont{Mendoza-Coto,
  Stariolo, and Nicolao}}]{Mendoza-Coto_PRL_2015}
\bibinfo{author}{\bibfnamefont{A.}~\bibnamefont{Mendoza-Coto}},
  \bibinfo{author}{\bibfnamefont{D.~A.} \bibnamefont{Stariolo}},
  \bibnamefont{and} \bibinfo{author}{\bibfnamefont{L.}~\bibnamefont{Nicolao}},
  \bibinfo{journal}{Phys. Rev. Lett.} \textbf{\bibinfo{volume}{114}},
  \bibinfo{pages}{116101} (\bibinfo{year}{2015}).

\bibitem[{\citenamefont{Schmalian and Wolynes}(2000)}]{Schmalian_2000_PRL}
\bibinfo{author}{\bibfnamefont{J.}~\bibnamefont{Schmalian}} \bibnamefont{and}
  \bibinfo{author}{\bibfnamefont{P.~G.}\bibnamefont{Wolynes}},
  \bibinfo{journal}{Phys. Rev. Lett.} \textbf{\bibinfo{volume}{85}},
  \bibinfo{pages}{836} (\bibinfo{year}{2000}).

\bibitem[{\citenamefont{Nussinov}(2004)}]{Nussinov_PRB_2004}
\bibinfo{author}{\bibfnamefont{Z.}~\bibnamefont{Nussinov}},
  \bibinfo{journal}{Phys. Rev. B} \textbf{\bibinfo{volume}{69}},
  \bibinfo{pages}{014208} (\bibinfo{year}{2004}).

\bibitem[{\citenamefont{Principi and Katsnelson}(2016)}]{Principi_PRL_2016}
\bibinfo{author}{\bibfnamefont{A.}~\bibnamefont{Principi}} \bibnamefont{and}
  \bibinfo{author}{\bibfnamefont{M.~I.} \bibnamefont{Katsnelson}},
  \bibinfo{journal}{Phys. Rev. Lett.} \textbf{\bibinfo{volume}{117}},
  \bibinfo{pages}{137201} (\bibinfo{year}{2016}).

\bibitem[{\citenamefont{Saratz et~al.}(2010)\citenamefont{Saratz,
  Lichtenberger, Portmann, Ramsperger, Vindigni, and Pescia}}]{Saratz_PRL2010}
\bibinfo{author}{\bibfnamefont{N.}~\bibnamefont{Saratz}},
  \bibinfo{author}{\bibfnamefont{A.}~\bibnamefont{Lichtenberger}},
  \bibinfo{author}{\bibfnamefont{O.}~\bibnamefont{Portmann}},
  \bibinfo{author}{\bibfnamefont{U.}~\bibnamefont{Ramsperger}},
  \bibinfo{author}{\bibfnamefont{A.}~\bibnamefont{Vindigni}}, \bibnamefont{and}
  \bibinfo{author}{\bibfnamefont{D.}~\bibnamefont{Pescia}},
  \bibinfo{journal}{Phys. Rev. Lett.} \textbf{\bibinfo{volume}{104}},
  \bibinfo{pages}{077203} (\bibinfo{year}{2010}).

\bibitem[{\citenamefont{Niemeijer and van Leeuwen}(1973)}]{NvL}
\bibinfo{author}{\bibfnamefont{T.}~\bibnamefont{Niemeijer}} \bibnamefont{and}
  \bibinfo{author}{\bibfnamefont{J.~M.~J.} \bibnamefont{van Leeuwen}},
  \bibinfo{journal}{Phys. Rev.} \textbf{\bibinfo{volume}{31}},
  \bibinfo{pages}{1411} (\bibinfo{year}{1973}).

\bibitem[{\citenamefont{Cannas}(1995)}]{Cannas1995}
\bibinfo{author}{\bibfnamefont{S.~A.} \bibnamefont{Cannas}},
  \bibinfo{journal}{Phys. Rev. B} \textbf{\bibinfo{volume}{52}},
  \bibinfo{pages}{3034} (\bibinfo{year}{1995}).

\bibitem[{\citenamefont{Delamotte}(2012)}]{Delamotte_Springer_2012}
\bibinfo{author}{\bibfnamefont{B.}~\bibnamefont{Delamotte}}, in
  \emph{\bibinfo{booktitle}{Renormalization Group and Effective Field Theory
  Approaches to Many-Body Systems,}}, edited by
  \bibinfo{editor}{\bibfnamefont{J.}~\bibnamefont{Polonyi}} \bibnamefont{and}
  \bibinfo{editor}{\bibfnamefont{A.}~\bibnamefont{Schwenk}}
  (\bibinfo{publisher}{Springer-Verlag}, \bibinfo{address}{Berlin Heidelberg},
  \bibinfo{year}{2012}).

\bibitem[{\citenamefont{{Giuliani} et~al.}(2011)\citenamefont{{Giuliani},
  {Lebowitz}, and {Lieb}}}]{Giuliani_PRB_11}
\bibinfo{author}{\bibfnamefont{A.}~\bibnamefont{{Giuliani}}},
  \bibinfo{author}{\bibfnamefont{J.~L.} \bibnamefont{{Lebowitz}}},
  \bibnamefont{and} \bibinfo{author}{\bibfnamefont{E.~H.}
  \bibnamefont{{Lieb}}}, \bibinfo{journal}{Phys. Rev. B}
  \textbf{\bibinfo{volume}{84}}, \bibinfo{pages}{064205}
  (\bibinfo{year}{2011}).

\bibitem[{\citenamefont{{Giuliani} et~al.}(2007)\citenamefont{{Giuliani},
  {Lebowitz}, and {Lieb}}}]{Giuliani_PRB_07}
\bibinfo{author}{\bibfnamefont{A.}~\bibnamefont{{Giuliani}}},
  \bibinfo{author}{\bibfnamefont{J.~L.} \bibnamefont{{Lebowitz}}},
  \bibnamefont{and} \bibinfo{author}{\bibfnamefont{E.~H.}
  \bibnamefont{{Lieb}}}, \bibinfo{journal}{Phys. Rev. B}
  \textbf{\bibinfo{volume}{76}}, \bibinfo{pages}{184426}
  (\bibinfo{year}{2007}).

\bibitem[{\citenamefont{Grousson et~al.}(2000)\citenamefont{Grousson, Tarjus,
  and Viot}}]{Viot}
\bibinfo{author}{\bibfnamefont{M.}~\bibnamefont{Grousson}},
  \bibinfo{author}{\bibfnamefont{G.}~\bibnamefont{Tarjus}}, \bibnamefont{and}
  \bibinfo{author}{\bibfnamefont{P.}~\bibnamefont{Viot}},
  \bibinfo{journal}{Phys. Rev. E} \textbf{\bibinfo{volume}{62}},
  \bibinfo{pages}{7781} (\bibinfo{year}{2000}).

\bibitem[{\citenamefont{Tranquada et~al.}(1995)\citenamefont{Tranquada,
  Sternlieb, Axe, Nakamura, and Uchida}}]{Tranquada_Nature_1995}
\bibinfo{author}{\bibfnamefont{J.~M.} \bibnamefont{Tranquada}},
  \bibinfo{author}{\bibfnamefont{B.~J.} \bibnamefont{Sternlieb}},
  \bibinfo{author}{\bibfnamefont{J.~D.} \bibnamefont{Axe}},
  \bibinfo{author}{\bibfnamefont{Y.}~\bibnamefont{Nakamura}}, \bibnamefont{and}
  \bibinfo{author}{\bibfnamefont{S.}~\bibnamefont{Uchida}},
  \bibinfo{journal}{Nature} \textbf{\bibinfo{volume}{375}},
  \bibinfo{pages}{561} (\bibinfo{year}{1995}).

\bibitem[{\citenamefont{Emery and Kivelson}(1993)}]{Emery_PhysC_1993}
\bibinfo{author}{\bibfnamefont{V.}~\bibnamefont{Emery}} \bibnamefont{and}
  \bibinfo{author}{\bibfnamefont{S.}~\bibnamefont{Kivelson}},
  \bibinfo{journal}{Physica C} \textbf{\bibinfo{volume}{209}},
  \bibinfo{pages}{597 } (\bibinfo{year}{1993}), ISSN \bibinfo{issn}{0921-4534}.

\bibitem[{\citenamefont{Keimer et~al.}(2015)\citenamefont{Keimer, Kivelson,
  Norman, Uchida, and Zaanen}}]{Keimer_Nature_2015}
\bibinfo{author}{\bibfnamefont{B.}~\bibnamefont{Keimer}},
  \bibinfo{author}{\bibfnamefont{S.~A.} \bibnamefont{Kivelson}},
  \bibinfo{author}{\bibfnamefont{M.~R.} \bibnamefont{Norman}},
  \bibinfo{author}{\bibfnamefont{S.}~\bibnamefont{Uchida}}, \bibnamefont{and}
  \bibinfo{author}{\bibfnamefont{J.}~\bibnamefont{Zaanen}},
  \bibinfo{journal}{Nature} \textbf{\bibinfo{volume}{518}},
  \bibinfo{pages}{179} (\bibinfo{year}{2015}).

\end{thebibliography}

\end{document}